\documentclass[%
 amsmath,amssymb,
 aps,
]{revtex4-2}

\usepackage{graphicx}% Include figure files
\usepackage{dcolumn}% Align table columns on decimal point
\usepackage{bm}% bold math
\usepackage{hyperref}% add hypertext capabilities
\hypersetup{
    colorlinks=true,
    linkcolor=blue,
    filecolor=magenta,      
    urlcolor=cyan,
    pdftitle={Overleaf Example},
    pdfpagemode=FullScreen,
    }

\begin{document}

\preprint{APS/123-QED}

\title{Controlling Vortex Rotation in Dry  Active Matter}% Force line breaks with \\
%\thanks{A footnote to the article title}%

\author{Felipe P. S. Júnior\(^{1}\)\footnote{felipe.junior@icen.br.ufpa},  Jorge L. C. Domingos\(^{2}\), W. P.
	Ferreira\(^{3}\), F. Q. Potiguar\(^{1}\)}
	%\thanks{A footnote to the article title}
 %\email{felipe.junior@icen.ufpa.br}
\affiliation{%
	\(^{1}\)Universidade Federal do Pará, Faculdade de Física, ICEN, Av. Augusto Correa, 1,Guamá, Belém, 66075-110, Pará , Brazil}%
	\affiliation{%
	\(^{2}\)MMML lab, Department of Physics, University of Latvia, Jelgavas 3, Rīga, LV-1004, Latvia}%
	\affiliation{%
		\(^{3}\)Universidade Federal do Ceará, Departamento de Física, Caixa Postal 6030, Campus do Pici, Fortaleza, 60455-760, Ceará, Brazil}%
	
	%MMML lab, Department of Physics, University of Latvia, Jelgavas 3, Rīga, LV-1004, Latvia
% \altaffiliation[Also at ]{Physics Department, XYZ University.}%Lines break automatically or can be forced with \\
%\author{ Jorge L. C. Domingos}%
 %\email{Second.Author@institution.edu}

%\collaboration{MUSO Collaboration}%\noaffiliation

%\author{Charlie Author}
 %\homepage{http://www.Second.institution.edu/~Charlie.Author}
%\affiliation{
 %Second institution and/or address\\
 %This line break forced% with \\
%}%
%\affiliation{
 %Third institution, the second for Charlie Author
%}%

%\collaboration{CLEO Collaboration}%\noaffiliation

\date{\today}% It is always \today, today,
             %  but any date may be explicitly specified

\begin{abstract}
We investigate the rotation of a vortex around a circular obstacle  in dry active matter in the presence of $M$ half-circles distributed around the obstacle. To quantify this effect, we define the parameter $\Pi_{M}$, which is the ratio between the mean angular velocity of the controlled vortex and the  root-mean-square angular velocity of the isolated vortex. We identify two rotational regimes determined by the obstacle configuration. In the first regime, where $\Pi_{M} < 0$ corresponding to the flat side of the half-circles facing the vortex, the rotation is clockwise. In the second regime ($\Pi_{M}>0$), it corresponding to the curved sides facing the vortex, the rotation becomes counterclockwise. 
We further analyze  the impact of this control on vortex stability, showing that   
the configuration of semi-circles can enhance or suppress stability depending on
their geometry and distance from the central obstacle. Our results demonstrate a possible setup to control the spontaneous rotation of dry active matter around circular obstacles.
\begin{description}
\item[Keywords]
Vortex, dry active matter, simulation, control.
\end{description}
\end{abstract}

%\keywords{Suggested keywords}%Use showkeys class option if keyword
                              %display desired
\maketitle

%\tableofcontents

\section{Introduction}
\label{sec:level1} 
In active matter, particles generate motion either by consuming their
internal energy or by extracting  energy from the environment \cite{marchetti2013hydrodynamics,das2020introductionSoftMatter}. A
collection of active particles is typically described using two distinct models:
dry models, in which particles do not move under the influence of hydrodynamic interactions, and wet models, where hydrodynamic interactions are taken into account. We can also consider dry models, where momentum is not conserved at the particle level. In these models, particles move
in a fluid that does not affect their motions, apart from viscous friction,
resulting in overdamped behavior. In contrast, wet models involve strong
fluid-particle interactions and thus, the total momentum (of both
particles and fluid) is conserved \cite{marchetti2013hydrodynamics}. Dry models can still be divided according
to the particle-particle interaction, which can either include particle alignment interactions, e.g., Vicsek model \cite{vicsek1995novel}	, or not include such interactions,
as Angular Brownian Motion (ABM) \cite{PhysRevLett.ABM2012,fily2014freezing,digrego2018}, which has a close relationship to
Run-and-Tumble dynamics \cite{cates2013active}.
%Active Matter is characterized by self-propelled particles (SSPs), such particles generate their movement by consuming internal energy or by using the energy present in the medium. We can simulate such particles through 2 Physical models:  flocking (Vicsek model \cite{vicsek1995novel}) and
%angular Brownian motion (ABM) types \cite{PhysRevLett.ABM,PhysRevLett.99.048102Run} which also includes run-and-tumble dynamics—RTD \cite{cates2013active}.
Active systems are known for their rich and intriguing phenomena far from equilibrium,
the spontaneous emergence of orientational motion order \cite{chate2006simple,vicsek1995novel},
motility-induced phase separation (MIPS)  in the absence of attractive forces \cite{PhysRevLett.ABM2012,CoarseningBeyondRipening_caporusso2023,redner2013,buttinoni2013dynamical,stenhammar2015activity,wittmann2016active,liebchen2017collective,caprini2020spontaneous,shi2020self,dittrich2021critical,rybak2022chiral},
motion rectification of active particles in asymmetric environment \cite{Rectification_Reichhardt2017,PhysRevE.Potigura2_2014,reichhardt2018clogging,PotiguarRevE2020}. 

%The clogging effect of self propelled particle in disordered media \cite{reichhardt2018clogging}. The active particles can  trapped due asymmetrical obstacles environment \cite{PotiguarRevE2020}.

The phenomenon of vortex formation in active matter has attracted significant research  in recent years.
It was shown experimentally that an active particle can be captured into closed orbits around a circular obstacle \cite{takagi2014hydrodynamic}; this occurs due to the relationship between the persistence time and the radius of curvature \cite{fily2014dynamics,pan2020vortex}. Motivated by the experimental work of Takagi \textit{et al.} \cite{takagi2014hydrodynamic}, Spagnolie \textit{et al.} \cite{spagnolie2015geometric}  showed that a swimmer approaching a small colloid is merely scattered. However, when the colloid exceeds a critical size, the swimmer becomes hydrodynamically captured. Wioland  \textit{et al.} \cite{wioland2016ferromagnetic} showed that 
hydrodynamically coupled vortex lattices can exhibit ferro- and antiferromagnetic-like
synchronization. Mokhtari  \textit{et al.} \cite{mokhtari2017collective} explained capture-and-release mechanism through which a vortex persists for extended periods, and reported that  obstacles often act as nucleation sites for particle accumulation and crystallization. Later, Pan \textit{et al.} \cite{pan2020vortex} showed a deep study in dry active matter of the vortex around a circular obstacle and they identified  three distinct regimes: random, transitional, and vortex. One year later, B. Qian \textit{et al.} \cite{qian2021absorption} reported an apparent rotation of a large aggregation of self propelled particles around lattices of tiny obstacles. Recently, in the context of wet models, Reinken \textit{et al.} \cite{reinken2022ising} demonstrated, using a continuum-theoretical approach, a nonequilibrium order-disorder transition in vortex lattices. In these lattices, vortices form at the center of unit cells defined by circular obstacles at their vertices. This transition is characterized as a second-order phase transition, with critical exponents consistent with those of the 2D Ising model. 
% including emergent structures with collective behavior distinguishing from that of the individual constituents. The study about collective behaviour of SSPs has revealed many nonequilibrium proprieties such as,
%the spontaneous appearance of motion orientational order \cite{chate2006simple,vicsek1995novel}, Phase Separation of SSPs  in the absence of attractive forces \cite{PhysRevLett.ABM2012} and  giant number fluctuations \cite{PhysRevLett.ABM,chate2006simple,ramaswamy2003active}. 
In all these investigations,  with wet models, there is a more extensive body of literature addressing vortex formation, which is consequence of mesoscale turbulence \cite{takagi2014hydrodynamic,wioland2016ferromagnetic,reinken2020organizing,reinken2022ising}. 
%while, the understanding of  vortex interactions in dry active models are still lacking.  

In dry active matter, the particles tend to accumulate around a circular obstacle; an imbalance between clockwise and counter-clockwise movers then gives rise to a spontaneous vortex \cite{mokhtari2017collective}; and the direction of rotational motion is randomly chosen in the system. Recently \cite{junior2024correlations} , we studied the interactions between two neighboring vortices and observed that 
 for an intermediate spacing between the obstacles, one vortex is formed by particles that aggregate around one of the obstacles, and, as a consequence, a non-vanishing particle current appears in the space between the obstacles. This current interacts with the particles that
 aggregate around the other obstacle, but do not form a vortex, 
 and drive them in a rotating sense that is contrary to the first 
 one. In terms of correlation, this case corresponds to negative correlation between these two vortices.
  But the direction of the induced vortex is randomly determined because the rotation of the first vortex  is also random.
   Hence, a natural question that arises from these observations is whether we can control the vortex rotation direction. Our results indicates this is possibility.
   Here, we demonstrate that it is possible to control the direction of vortex rotation introducing half-circles obstacles around the central circular obstacle.
   The distance between the obstacle and the half-circles, and the orientation of the latter relative to the principal directions of the system, mainly determine the rotation direction of the vortex.

   %which guide the particles near the circular obstacle. The vortex rotates clockwise when the flat sides of the semi-disks face the central disk, and counterclockwise when the curved sides of the semi-disks are oriented toward the central disk.

This manuscript is organized as follows. Our model system
is presented in Sec. \ref{Model}. The numerical results and discus-
sion are presented in Sec. \ref{results}. Our conclusions are given in
Sec. \ref{conclusions}.
\section{Model}
\label{Model}
\begin{figure}[ht]%%%%%%%%Figure 2
 	\centering
 	\includegraphics[width=0.80\linewidth]{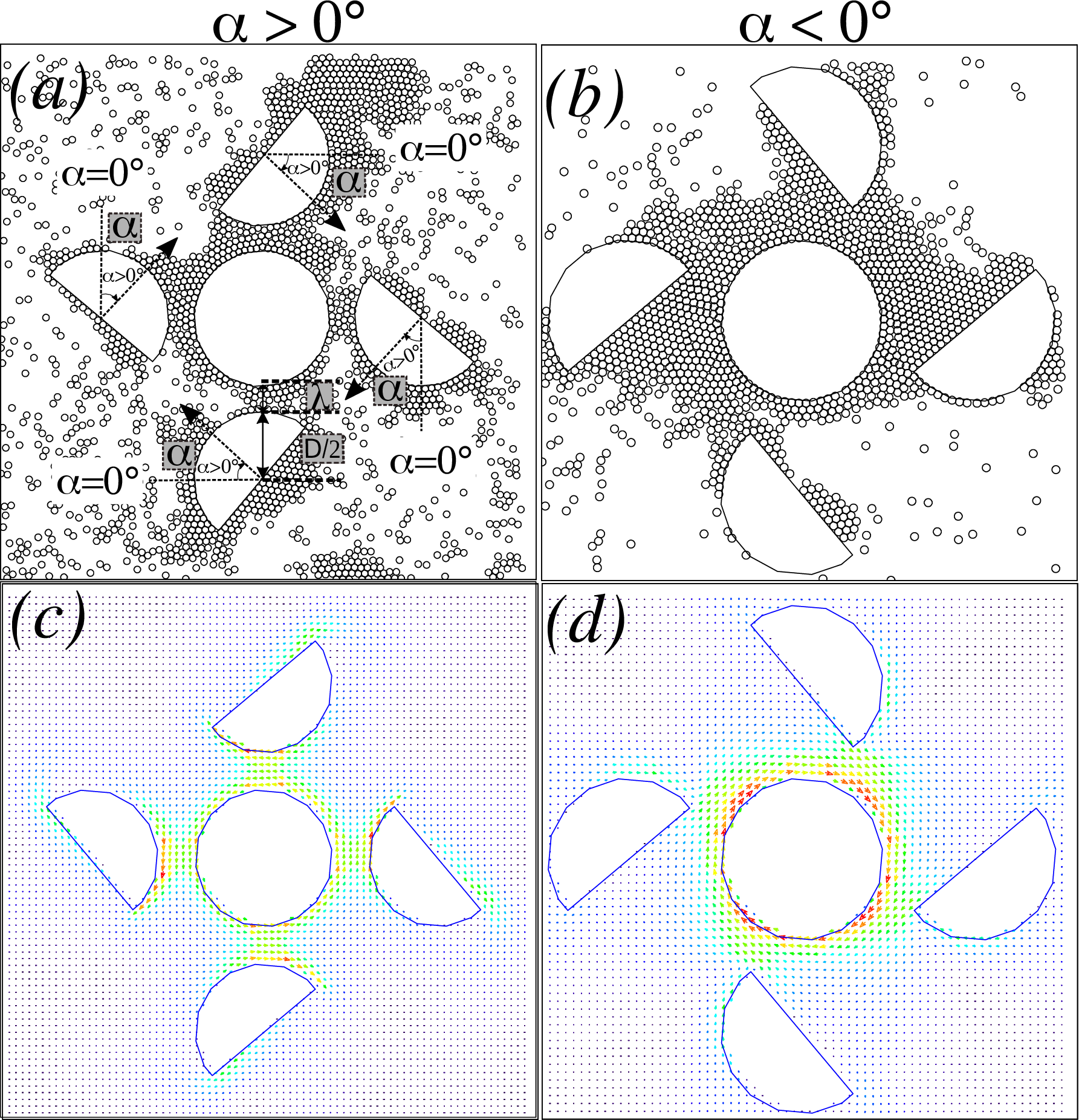}
 	%The mean velocity field is derived from averaging 25 long-time-averaged flow fields depicted in Figs. (a) and (b). .
 	\caption{ In the upper panels, two snapshots of  the system configuration with obstacle size $D=20$ and $M=4$  half-circles, for:  (a)   $\alpha=50$, $\lambda=0.3D$ and (b)  $\alpha=-50$, $\lambda=0.1D$. In lower panels, corresponding mean velocity fields averaged over 25 long-time realizations are presented in (c) and (d).}
 	\label{FIG1}
 \end{figure}

Our model consists of a two-dimensional (2D) system of
$N$ soft active particles, with diameter $\sigma$ inside a square box of side $L$. A central circular obstacle, with diameter $D$, surrounded by $M\in \{1,2,3,4\}$ half-circles, also of diameter $D$, with their centers located at the fixed positions $(L/2,L/2\pm(\lambda+D/2))$ and $(L/2\pm(\lambda+D/2),L/2)$. We define $\lambda$  as  the smallest distance between the curved side of the half-circles and the surface of the central circular obstacle, {\em i.e.}, the gap between the half-circles and the central obstacle. We set the box size as $L=3(D+\lambda)$. 
 We rotate each half-circle by the angle $\alpha$ around the system's principal direction, which induces particle currents along the normals of their flat sides \cite{PhysRevE.Potigura2_2014}: The positive angles, $\alpha > 0$, correspond to rotations of the half-circles where their curved sides face the circular obstacle, while for negative angles, $\alpha<0$, correspond to rotation of the half-circles where their flat sides face the central obstacle. see Fig. \ref{FIG1}(a).

%2\begin{figure}[ht]
    %\centering
    %\includegraphics[width=0.8\linewidth]{Fig1.eps}
    %\caption{Caption}
    %\label{model_angle}
%\end{figure}
The particles interact through a  linear spring force law, $\textbf{F}_{ij}=\kappa(d_{ij}-r_{ij})\hat{\textbf{r}}_{ij}$, for $r_{ij}< d_{ij}$ (otherwise $\textbf{F}_{ij}=0$), with $i\neq j$. Here,  $r_{ij}=|\textbf{r}_{i}-\textbf{r}_{j}|$ is the distance between particles, and $d_{ij} = (d_{i} + d_{j} )/2$ is the mean contact diameter. For a particle-particle contacts, $d_{ij}=\sigma$; for particle-obstacle contacts, $d_{ij}=(\sigma+D)/2$; and for  contacts with a flat side of the half-circle, $d_{ij}=\sigma/2$.

The motion of the $i$-th active disk is described by the
following equations:
\begin{equation}
    \textbf{v}_{i}=\textbf{v}_{i}^{f}+\mu\textbf{F}_{i}+ \mu\boldsymbol{\xi}_{i}(t), 
\end{equation}
 
\begin{equation}
\frac{d\theta_{i}}{dt}=\eta_{i}(t),
\end{equation}
where $\mu$ is the motility, $\textbf{F}_{i}=\sum_{j}\textbf{F}_{ij}$, 
$\textbf{v}^{f}_{i}=v_{0}(\cos\theta_{i}\hat{x}+\sin\theta_{i}\hat{y})$,
is the active velocity, of magnitude $v_{0}$, and random direction $\theta_{i}$;  $\boldsymbol{\xi}_{i}(t)$ is a random thermal force and $\eta_{i}(t)$ is the random angular velocity. In all simulations we employ periodic boundary conditions (PBC) along x, y directions. Both variables are Gaussian white noises that follow the rules $\langle\boldsymbol{\xi}_{i}\rangle=0$ and
$\langle\xi_{i\alpha}(t)\xi_{j\beta}(t^{\prime})\rangle=2\xi\delta_{ij}\delta_{\alpha\beta}\delta(t-t^{\prime})$, where $\alpha,\beta=x,y$ and
$i,j\in [1,N]$; $\langle\eta_{i}(t)\rangle=0$ and $\langle\eta_{i}(t)\eta_{j}(t^{\prime})\rangle=2\eta\delta_{ij}\delta(t-t^{\prime})$, $\xi$ and $\eta$ are the noise intensities; we consider athermal particles, \textit{i.e}
%%%%>>>>>>>>>>>>Talves mude isso quando fazer com diametro diferente de 20
$\xi=0$. The other parameters are set as	: $\sigma=1$, $v_{0}=1$, $\mu=1$,
 $\kappa=50$ and $\eta=0.001$ (for a particle-obstacle contact $\kappa_{\mathrm{obs}}=1000$).

 We calculate the vortex angular velocity $\omega$ taking into account only particles that are within distance $r_{\mathrm{b}}$ of  the central obstacle. The expression reads, for an instant $t$:
 \begin{equation}
 \omega(t)=\frac{1}{N_{b}}\displaystyle\sum_{
\begin{array}{c}
r_{ic}<r_{b}
\end{array}
} \Big(\frac{\textbf{v}_{i}\cdot\hat{\boldsymbol{\psi}}_{i}}{r_{ic}}\Big)\label{eq:omega}
 \end{equation}
 where, $r_{ic}=|\textbf{r}_{i}-\textbf{r}_{c}|$ with ${\bf r}_{\mathrm{c}}=L/2(\hat{\bf i}+\hat{\bf j})$ the position of the central obstacle, and $\hat{\boldsymbol{\psi}}_{i}$ is the unit tangential vector in the polar direction and $N_{b}$ is the number of particles that satisfy $r_{i\mathrm{c}}<r_{\mathrm{b}}$. 
We observe that the angular velocity remains approximately constant
 for $r_{b} < D/2 + 4\sigma$.	 
 We choose $r_{b}=D/2+2\sigma$. From (\ref{eq:omega}), we define the parameter
 \begin{equation}
     \Pi_{M}=\frac{\Big\langle\omega_{M}(t)\Big\rangle}{\sqrt{\Big\langle\omega_{0}^{2}(t)\Big\rangle}}\label{eq:orderparameter}
 \end{equation}
 where $\langle ...\rangle$ denotes a mean over time and distinct realizations, $\omega_{M}(t)$ is angular velocity  of the vortex around the circular obstacle surround by $M$ half-circles  and $\omega_{0}(t)$ is angular velocity of the isolated vortex (when both, $M$ and $\lambda$ are zero).
 This parameter quantifies the average angular velocity of the controlled vortex relative to that of an isolated vortex. Note that, since the mean angular velocity of an isolated vortex vanishes, the denominator in Eq. (\ref{eq:orderparameter}) corresponds to root mean square (RMS) of the angular velocity of the isolated vortex.
In our simulations, we considered $\alpha\in[-90^{\circ},90^{\circ}]$ with increments of $\Delta\alpha=10$, $D\in[10, 40]$ and $\lambda\in[0.2D, 1D]$ for all diameter sizes.  
Additionally, for $D = 20$, we also considered $\lambda\in [0.1D, 0.3D]$.
Our numerical investigation was carried out for area fraction $\phi=\frac{1}{4}N\pi\sigma^{2}/(L^{2}-A_{obs})=0.3$, where $A_{obs}=\frac{D^{2}\pi}{4}(1+\frac{M}{2})$ is the area occupied by obstacles. Each run consisted of either $2 \times 10^6$ or $1 \times 10^7$ times steps, depending on the statistical accuracy required, with the thermalization steps set to  $1 \times 10^6$ and $2 \times 10^6$, respectively. The integration time step was set to $h = 10^{-3}$ in all cases. We performed our measurements over $25$ independent realizations of the initial conditions, and over $5$ realizations for cases involving longer simulation times.

 \section{results and discussion}
 \label{results} 
 We now present our results for  vortex rotation in the presence of the surrounding half-circles. We begin by addressing the behavior of  $\Pi_{M}$, Eq. (\Ref{eq:orderparameter}), as a function of the angle $\alpha$ and the gap, $\lambda$.
 \subsection{Control of vortex rotation}
 We first analyze qualitatively how the semi-circular obstacle influence the direction of vortex rotation.
As shown  in Ref. \cite{junior2024correlations}, when two vortices are a certain distance apart, and have negative correlations, one of them can drive the (usually opposite) rotation direction of the other. In Fig. \ref{FIG1},
we show theb mean velocity field $\Big\langle\sum_{i}\textbf{v}_{i}\delta(\textbf{r}_{i}-\textbf{r})\Big\rangle$  for both positive [Fig. \ref{FIG1} (c)] and negative [Fig. \ref{FIG1} (d)] values of $\alpha$ together with their corresponding configuration snapshots, [Figs. \ref{FIG1} (a) and \ref{FIG1}(b)]. We observe that the mean vortex rotation is  clockwise (counterclockwise) for negative (positive) values of $\alpha$. 
For $\alpha>0$, particle currents between the semi-circular obstacles and the central disk  develop along their normals to the flat sides, driving a counter-clockwise rotation (Movie 1).  In contrast, for $\alpha<0$, 
the particle currents induced by the half-circles, tend to move (Movie 2) particles along the clockwise direction \cite{rojas2023mixtures,PhysRevE.Potigura2_2014}. Hence, our setup allows control over the vortex direction through a mechanism similar to that described for two circular obstacles \cite{junior2024correlations}.

\begin{figure}[ht]
	\centering
	\includegraphics[width=0.75\linewidth]{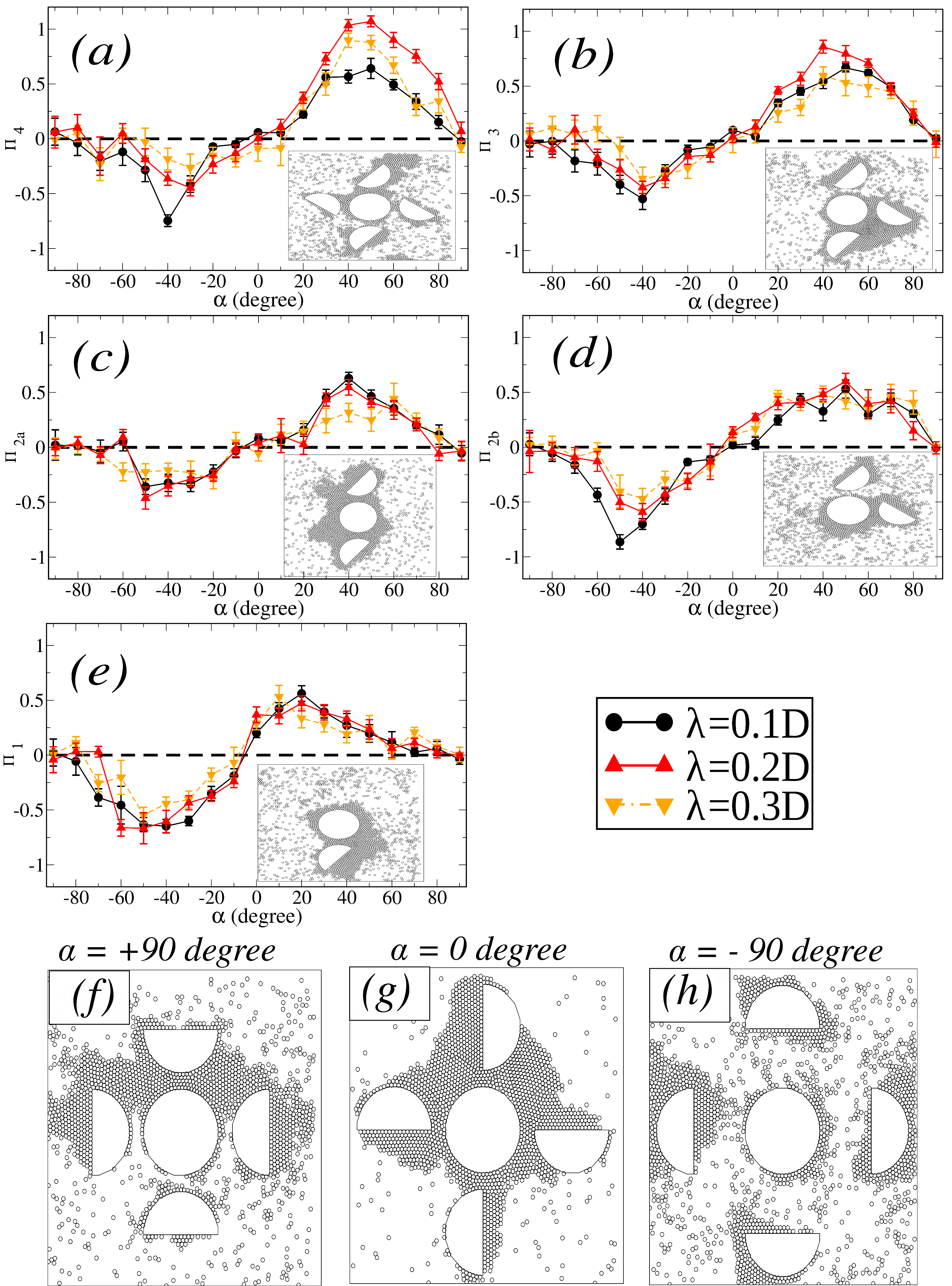}
	\caption{$\Pi$ as a function of the angle $\alpha$ for different values of $\lambda$. 
		Each panel shows the behavior of $\Pi$ for a distinct geometric configuration of half-circles surrounding the central obstacle, with the configuration displayed as an inset: 
		(a) four half-circles (4HC), (b) three half-circles (3HC), (c) two adjacent half-circles (2AHC), (d) two opposite half-circles (2OHC), and (e) one half-circle (1HC). 
		Panels (f)--(h) show snapshots of the system for an obstacle of size $D = 20$ with $M = 4$ half-circles at $\lambda = 0.2D$, for $\alpha = +90^\circ$, $0^\circ$, and $-90^\circ$, respectively.}
	\label{FIG2}
\end{figure}

We also study the effect of the number of half-circles on the rotation of the central vortex.
In Fig. \ref{FIG2}, we show $\Pi_{M}$ as a function of $\alpha$ for obstacle size $D=20$ and  different values of the parameter $\lambda$, for distinct arrangements of half-circles: four half-circles (4HC) [Fig. \ref{FIG2}(a)]; three half-circles (3HC) [Fig. \ref{FIG2}(b)]; two adjacent half-circles (2AHC) [Fig. \ref{FIG2}(c)]; two opposite half-circles (2OHC) [Fig. \ref{FIG2}(d)]; and one half-circle (1HC), [Fig. \ref{FIG2}(e)].
The function $\Pi_{M}(\alpha)$ exhibits two distinct regimes in all cases. The first regime, where $\Pi_{M} < 0$, occurs for $ -90^{\mathrm{o}}  < \alpha< 0$, corresponding to the flat sides of the half-circles facing the vortex. The second regime, where $\Pi_{M}>0$, occurs for $0 < \alpha < 90^{\mathrm{o}}$, where the curved side of the half-circles face the vortex, indicating 	counterclockwise 
rotation. The curves for positive $\alpha$ are also less noisy and closer to each other compared to the ones for negative $\alpha$. We attribute this difference to the fact that, for $\alpha<0$, the gap $\lambda$ depends on $\alpha$. Still, we keep the notation for a fixed $\lambda$ in these cases since our focus is on the qualitative features of the control with the flat sides. 
In other words, we sought to confirm our reasoning that the particle currents induced by the half-circles near the vortex determine the vortex rotation direction, 
though a more detailed study is needed for $\alpha<0$ to obtain more quantitative results.
Finally, a clockwise rotation could also be obtained, if the half-circles are rotated  from $90^{\mathrm{o}}$ to $180^{\mathrm{o}}$, since the particle flow at each half-circle would invert relative to that obtained at
$90^\mathrm{o}$. Similarly, a counterclockwise rotation can be obtained if we rotate the half-circles by $-180^\mathrm{o}<\alpha<-90^\mathrm{o}$.

%%% keep that paragrapho
%In Fig. \ref{FIG2}(a), for the 4HC case, the onset of the negative regime of the $\Pi_{4}$ function is around $\alpha = -70$. This indicates the existence of a minimum rotation angle of the half-circles' flat side necessary to obtain a non-zero value of the $\Pi$ function. Specifically, at $\alpha = -90$, there is no significant interaction between the particles at the surfaces of the half-circles and vortex particles, which prevents the controlling of the particle vortex motion. From $\alpha = -70$ to $-20$, we observe a negative valley in the $\Pi_{4}$ function, indicating a clockwise-controlled vortex. The  minimum value of that valley, $ \Pi_{4,\min}(\lambda)$,  depends on the parameter $\lambda$. For example, for $\lambda = 2$, the minimum value is approximately $\Pi_{\min} \approx -0.75$, which is lower than for $\lambda=6$, $\Pi_{\min}\approx -0.25$. 
%This occurs because, as the parameter $\lambda$ increases, the central vortex moves farther away from the boundary of the semi-disks.

We now focus on the 4HC arrangement [Fig. \ref{FIG2}(a)]. For positive $\alpha$, when the distance between the central obstacle and curved surfaces of half-circles does not change with angle $\alpha$, we 
find that $\Pi_{4}=0$ for $\alpha=0$, and  reaches a peak at $\alpha\approx 50^{\mathrm{o}}$ for most of the $\lambda$ curves. At this peak, $\Pi_{4}$ is slightly greater than $1$ for $\lambda=0.2D$, indicating that the controlled vortex may rotate faster than the isolated one. Beyond this peak, $\Pi$ decreases monotonically,  vanishing at $\alpha = 90^{\mathrm{o}}$.
A similar behavior is observed for negative $\alpha$ with $\Pi_{4}$ eventually vanishing at $\alpha=-90^\mathrm{o}$.

 This behavior is consistent with our setup: at $\alpha=90^{\mathrm{o}}$ [Fig. \ref{FIG2}(f) and \ref{FIG2}(h)],
 the particle currents induced by the half-circles closest to the obstacle vanish, since the half-circles push particles symmetrically in opposite, horizontal (top and bottom) and vertical (left and right).
 At the other boundary, $\alpha = 0$ [Fig. \ref{FIG2}(g)], there is a change-over point in for the particle current directions: as $\alpha$ increases or decreases, either curved  or the flat sides face the obstacle, and because they induce currents in opposite directions, the net current vanishes. Consequently,
    the rotation reaches maximum, around the middle of the interval at $\alpha \approx45^{\mathrm{o}}$, as expected from the obstacle geometry. It is also clear from this scenario that the vortex rotation in the range $\alpha\in[90^{\mathrm{o}},180^{\mathrm{o}}]$ should be opposite to that observed in $\alpha\in[0,90^{\mathrm{o}}]$, and should vanish again, at $\alpha=\pm180^\mathrm{o}$. The influence of $\lambda$ is not very strong:  $\Pi_{4}$ decreases slightly
 around the peak only for $\lambda=2$, compared with the curves at larger $\lambda$.
 
 % when lambda increases. This corroborates the expectation that closer half-circles would have a larger influence on the vortex rotation than those located farther away. 

 %In Fig.~3(a), the maximum of the $\Pi$ function is achieved for $\alpha < 10$. For example, when $\alpha$ varies from 30 to 60, corresponding to $\lambda = 6$, $\Pi$ remains approximately constant at its maximum value, $\Pi \approx 1$. Beyond this range, $\Pi$ decreases monotonically, reaching zero at $\alpha = 90$. As the angle $\alpha$ increases, $\Pi$ rises and reaches its peak to $\alpha =[30,60] $. This maximum, denoted as $\Pi_{\text{max}} = \Pi_{\text{max}}(\lambda)$, depends on the value of $\lambda$. 
  %On the negative side ($\alpha < 0$), valleys are observed. As $\alpha$ decreases, $\Pi$ decreases until it reaches a minimum at $\alpha = -50$. This minimum value, $\Pi_{\text{min}}=\Pi_{\text{min}}(\lambda)$, is also dependent on the $\lambda$ parameter, for example for $\lambda=12$ the $\Pi_{\text{min}}\approx -0.9$  that is less than for $\lambda=16$ $\Pi_{\text{min}}\approx -0.25$. Beyond $\alpha = -50$, as $\alpha$ approaches -90, the $\Pi$ function gradually returns to zero.

%Both phases are influenced by the configuration of the half-disks and the parameter $\lambda$. 

% 2OSD, and 1SD 
We now consider the results of 
Figs.~\ref{FIG2}(b), (c), (d), and (e)---corresponding to the 3HC, 2AHC, 2OHC, and 1HC configurations, respectively. The amplitude of the $\Pi_{M}$ curves is lower compared to the 4HC case on both the flat and curved sides, but all curves are qualitatively similar to those of Fig. \ref{FIG2}(a), which reflect the fact that the half-circles are the cause of the directed vortex rotation. Finally, we observe that the maximum value of $\Pi_{M<4}$ is never greater than $1$, meaning that for $M<4$ the controlled vortex does not rotate as fast as an isolated one, in contrast to the result for $M=4$.
We can still see the maximum and minimum values of $\Pi_{M<4}$ for $\alpha>0$ and $\alpha<0$, but in some cases, the angles at which these extreme values occur are not $45^\mathrm{o}$. For instance, for 1HC, Fig. \Ref{FIG2}(e), $\Pi_{1,max}$ occurs between $\alpha=10^{\mathrm{o}}$ and $20^{\mathrm{o}}$. 
It can be observed that the curves on both $\alpha$ regimes (as seen in the 4HC case)are nearly  independent of the
 gap, which is rather surprising since we would expect the control of the vortex rotation to decrease as we consider farther half-circles.
 %is less organized than an isolated one.
 %%corrigir aqui Felipe jr
  In the next section, where we discuss the effect of the surrounding half-circles on the vortex stability regime \cite{pan2020vortex}.

 % $\lambda$ parameter. This is a evidence that with retired of half-circles   around the circular one, the control of the vortex is independent of the $\lambda$ parameter, although the control even exist with angle $\alpha$.

%In Figs.~\ref{FIG3}(b), (c), (d), and (e)---corresponding to the 3SD, 2ASD, 2OSD, and 1SD configurations, respectively---the amplitude of the $\Pi$ curves is lower compared to the 4SD case on both the flat and curved sides. The dependence of the maximum and minimum values on the $\lambda$ parameter diminishes as the semi-disks are removed, particularly on the curved side, where this effect is observed more rapidly. For example, in Fig.~\ref{FIG3}(b) (3SD configuration), the curves on the curved side ($\alpha>0^{\circ}$) display a noticeable loss of sensitivity to changes in $\lambda$, with all curves converging to nearly the same maximum $\Pi$ value. In contrast, on the flat side, significant variations on $\Pi$ curves are still observed with change of $\lambda$, where the valleys still at $\lambda=12$ and $\lambda=13$. In Fig.~\ref{FIG3}(e), it can be observed that the curves on both sides ($\alpha < 0$ and $\alpha > 0$) are independent of the $\lambda$ parameter. This is a evidence that with retired of semi disks   around the circular one, the control of the vortex is independent of the $\lambda$ parameter, although the control even exist with angle $\alpha$.

\subsection{The issue of the vortex stability}
\begin{figure}
	\centering
	\includegraphics[width=\linewidth]{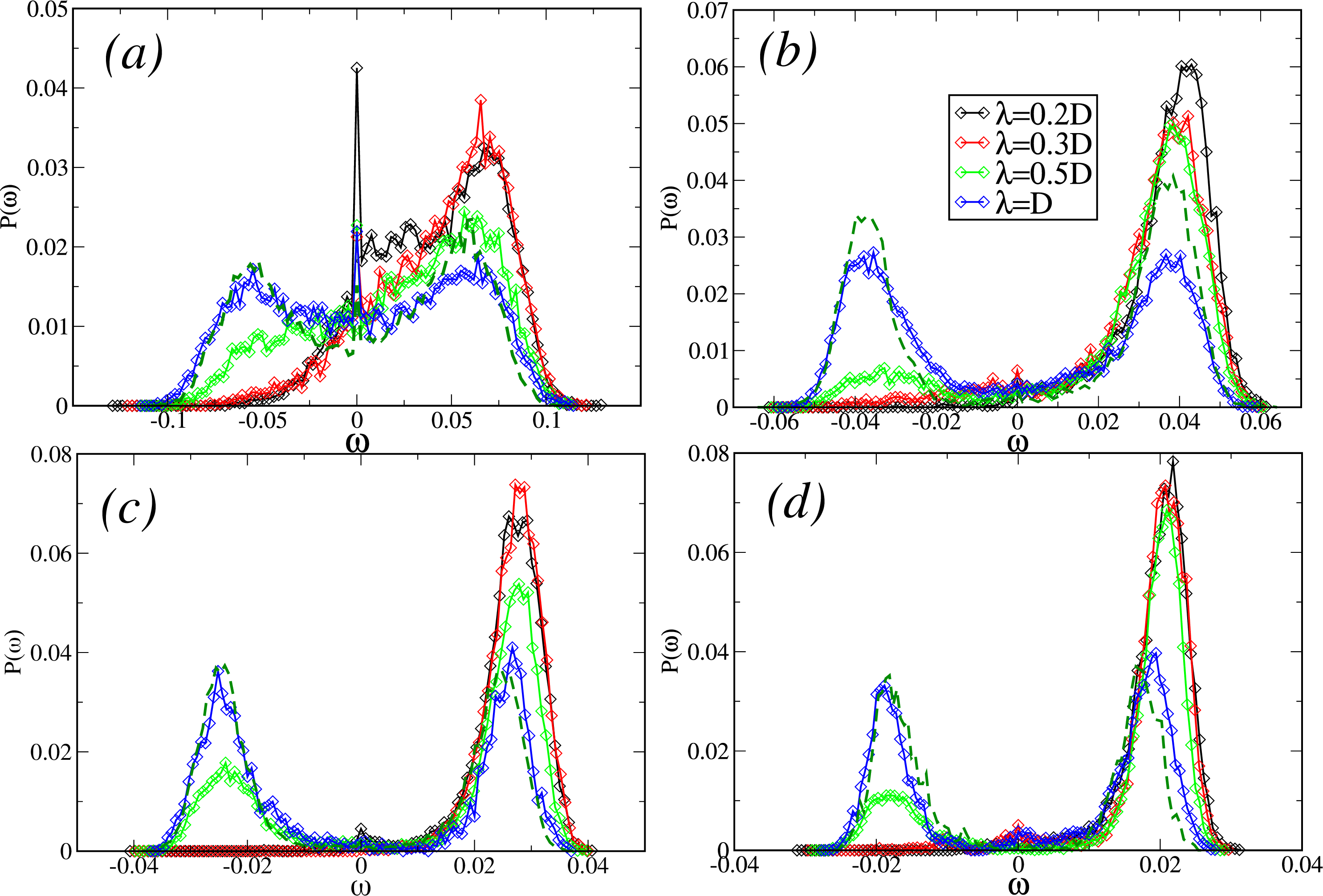}
	\caption{Normalized angular velocity probability distributions $P(\omega)$ for $\alpha = 50^\mathrm{o}$ in the 4HC for different $\lambda$ and different obstacle sizes: (a) $D = 10$, (b) $D = 20$, (c) $D = 30$, and (d) $D = 40$, and the dashed curves represent the isolated case, where box's size corresponds  $L = 3D$.}
	\label{FIG3}
\end{figure}

We now discuss the effect of the half-circles of the vortex stability, focusing on the 4HC case with $\alpha=50^{\mathrm{o}}$, which is around the maximum observed for $\Pi_{4}$. Pan \textit{et al.} \cite{pan2020vortex} showed
a strong dependence of vortex stability on the obstacle diameter, highlighting
the importance of this parameter in vortex formation.
It was shown that the vortex is 
unstable or random for $D\leq 10$ 
and swiftly oscillates between the two rotation directions, whereas for $20\leq D<40$ it is transient. For $D\geq 40$, the vortex is in a stable state. Hence, a natural question in the present system is how the half-circles affect these three regimes of vortex rotation.
 To gain further insight into the stationary state of the system, we present in Fig. \Ref{FIG3} the normalized probability distribution for the vortex angular velocity, $P(\omega)$, for $D=10$, $20$, $30$ and $40$; and four distinct values of  $\lambda=0.2D$, $0.3D$, $0.5D$ and $D$. For comparison, the angular velocity of an isolated vortex is also shown. For all values of $D$, the green curves (isolated vortex) are approximately bimodal 
    and nearly symmetrical with respect to the angular velocity, indicating that the isolated vortex has no preferential rotation direction, as expected. 
 For $D=10$, we see a peak at $\omega=0$,  indicating the 
  frequent changes of rotation in this regime.
 When the half-circles are introduced, the most evident effect  for all, for 
  $\lambda\leq 0.5D$, is a preferential counterclockwise vortex rotation direction,
   which reflects the data shown in Fig. \ref{FIG2}. For $\lambda=D$,
    the distribution returns to its bimodal shape, indicating that the vortex becomes, effectively, isolated; in other words, the effect of the half-circles decreases with their increasing distance to the central obstacle.
 Notice that this effect is not evident in the results of Fig. \ref{FIG2}, where $\lambda$ is relatively low.
 
 Another notable feature of the $P(\omega)$ distributions is that the right-hand peak velocities, $\omega^{*}$, remain approximately constant for a fixed obstacle size $D$. 
 For gaps $\lambda \leq 0.5D$, these peak values are slightly higher than those of the isolated vortex, indicating that the vortex rotates marginally faster when the semi-circular obstacles are placed closer to the central disk. 
 This behavior is consistent with the overall trend observed in the data. 
 Furthermore, $\omega^{*}$ decreases systematically with increasing $D$, reflecting the reduced angular velocity of larger vortices. 
 The influence of the control geometry on stability differs across the unstable, transient, and stable regimes. 
 We begin by examining the unstable regime in more detail.

 %The tails of the distributions also become smaller for larger obstacles; this is an effect of the overdamped regime, which leads to $\omega_{M} \sim \frac{2v_{0}}{D}$

In
Fig. \ref{FIG3} (a), ($D=10$) the vortex oscillates between both directions, as seen from the peak at $P(\omega=0)$ for all values of $\lambda$. Due to the half-circle setup, a competition arises between large fluctuations of the vortex and particle currents induced by the half-circles (Movie 3). These strong fluctuations tend to invert the rotation, but the half-circles continuously push particles towards the induced rotation. If we define a vortex as unstable when $P(\omega=0)$ is comparable to its peak value $P(\omega^{*})$, then for $D=10$ the half-circles do not change this feature, they only bias the vortex toward one direction when the strong fluctuations within allow it.

%Below, we will discuss a distinct condition for the vortex stability, we will see in more detail the influence of the half-circles on this feature.

 %, the pronounced peak of probability curve at $\omega=0$ in Fig. \ref{FIG3}. (a) to $\lambda=0.2D$ is evidence this mechanism. 
 %This tendency reappears repeatedly, and in this case, the preferred direction is positive (with $\alpha = 50^\circ$).
For more stable vortices ( $D=20$, $30$, and $40$), the normalized probability curves are smoother with sharper peaks than for $D=10$ and 
small $P(\omega=0)$. Note that there is still a finite probability to observe zero rotation, but it is rather small given the large obstacle sizes. Hence, the presence
of the half-circles does not alter the intrinsic stability of a vortex.

% Hence, the presence of the half-circles does not alter the intrinsic stability of a vortex.

%%%%
%%%FIGURE
\begin{figure}[ht]
	\centering
	\includegraphics[width=0.7\linewidth]{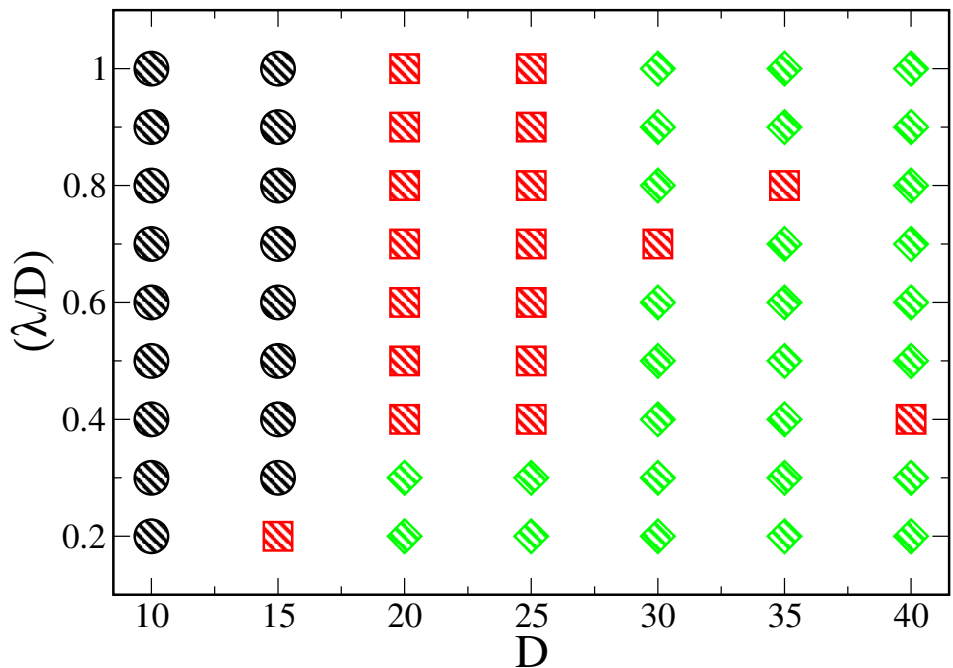}
	\caption{ Phase diagram of $\lambda/D$ vs $D$  with $\alpha=50^{\mathrm{o}}$ and 4SD. Black circles, red squares and green diamonds
		represent the unstable, transient, and stable states, respectively.}
	\label{FIG5}
\end{figure}
After describing the influence of the half-circles on the vortex stability, we propose a quantitative condition for the stability of a particular state. In the random and transient states, there is a probability that a vortex changes its rotation direction within the observation time. Our condition is based on counting such changes during the system's evolution. We divide a long simulation, with a total simulation time step of $10^{7}$, into time step windows of length $10^{6}$ and start counting the changes of direction at time step $2\times 10^{6}$.

\begin{figure}[ht]
		\centering	
	\includegraphics[width=\linewidth]{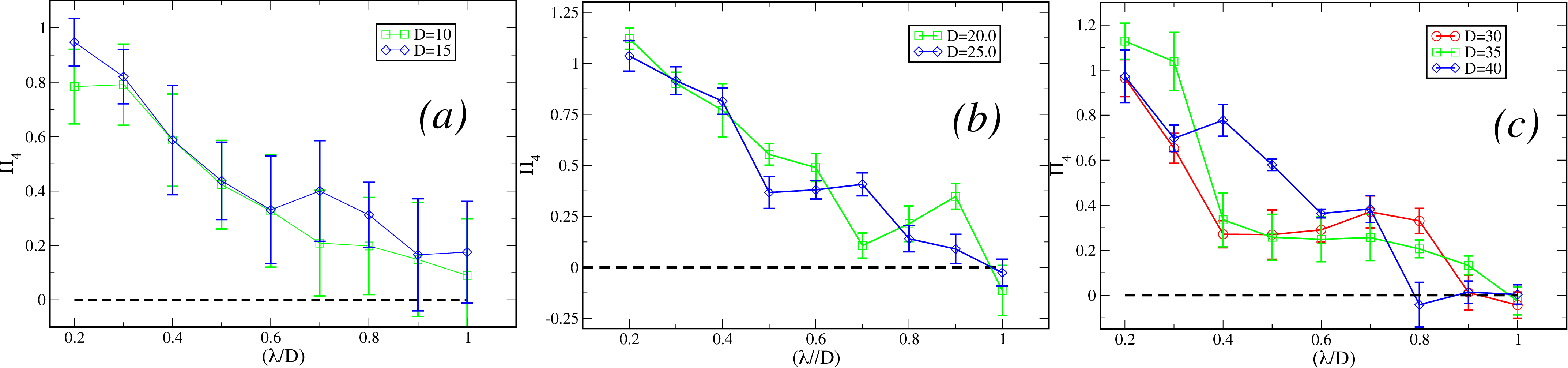}
		\caption{ Dependence of $\Pi_{M}$ on $\lambda$
			 for $\alpha = 50^{\mathrm{o}}$ in the 4HC. Panels (a), (b) and (c) corresponds to the naturally random,transient and stable vortex states respectively.}
	\label{FIG4}
\end{figure}

We only consider changes in the vortex rotation direction if the cumulative time intervals between reversals within a given window is at least $10^5$ steps.
 However, in the case of a single isolated reversal within a time window, we take the time interval from the reversal to the end of the window and assess whether it is at least $10^{5}$ steps, which is the threshold for a valid change.
 The analysis is performed over time windows of length $\delta\tau = 10^{6}$ steps. Whenever a window contains at least one point that satisfies this criterion, it is identified as an instability event. The frequency of instability events is defined as

\begin{equation}
	f=\Big\langle\frac{n_{ins}}{n_{win}}\Big\rangle_{r}\label{eq:frequency},
\end{equation}
where $n_{\text{ins}}$ is the number of time windows with instability events and $n_{\text{win}} =\frac{(10^{7}-2\times 10^{6})}{\delta\tau}= 8$.  The notation $\langle \dots \rangle_r$ denotes the average over the number of independent runs (eight realizations). For $f < 0.1$, the vortex is stable; for $0.1 \leq f < 0.5$, the vortex  is transient; and for $f \geq 0.5$, it is 
unstable. We show a phase diagram for $f$ in the $\lambda/D$ vs. $D$ space in Fig. \ref{FIG5}; black circles denote random vortex, red squares transient vortex, and green diamonds stable vortex.
%We verify that,the vortex state of the isolated vortex is similar to  the state of the non-isolated vortex  when $\lambda=D$ for all cases, and its results are omitted on diagram \ref{FIG5}.
As already mentioned, the influence of half-circle setup vanishes  as $\lambda\rightarrow D$, 
and the vortex becomes effectively isolated (compare the $\lambda=D$ line to Figs. \ref{FIG4}(a)-(c) \cite{pan2020vortex}).

%where the system return to an isolated vortex configuration, see Fig.\ref{FIG3}. $10 \leq D\leq 15$ corresponds to a random vortex state, $20\leq D\leq25$ correspond to a transient vortex state, and $30 \leq D\leq 40$ correspond  to a stable vortex state.

 The random vortex state appears only for $D\leq15$. For $D=10$, only unstable states are observed whereas at $D=15$, a transition to transient vortex state occurs for $\lambda/D\leq 0.2$. This shown that, besides controlling the vortex direction, the half-circles setup also affects vortex stability. Similar effect is also observed to naturally transient vortices, $D\geq20$. For $D=20$ and $D=25$, we observe  stable vortex states appear only for  $\lambda/D\leq0.3$, while the transient state occurs at $\lambda/D> 0.3$.

For $D>=30$, where the stable vortex state dominates, we see three transient states, $D=30$, $\lambda=0.7D$; $D=35$, $\lambda=0.8D$; and $D=40$, $\lambda=0.4D$. We verified that these cases are not due to poor statistic. Instead, they correspond to points where
the influence of the half-circles ceases to determine the stationary rotation. For large obstacles, the vortex naturally selects a rotation direction, which may or may not coincide with that imposed by the half-circles. When the latter are close to the obstacle, the vortex follows their influence; when they are far from the obstacle, it behaves essentially as an isolated stable vortex, which seldom changes the rotation direction. Hence, at certain parameter values, this regime changes, corresponding to the three red points. A similar change in the vortex stability is also observed for $D<30$: the system always evolves from more stable (under the influence of half-circles) to less stable (effectively isolated vortex).

We now present some results of the dependence of $\Pi_{4}$ on $\lambda/D$, over the full range of $\lambda$ studied, at  $\alpha = 50^{\mathrm{o}}$. Fig.~\ref{FIG4}(a) shows the case $D \leq 15$  (effectively random), Fig.~\ref{FIG4} shows $20.0\leq D\leq 25 $ (effectively transient) and Fig.~\ref{FIG4}. (c) shows $ D\geq 30$ (effectively stable). In all cases, the maximum value of $\Pi_{4}$ function occurs at $\lambda=0.2D$ and decreases with $\lambda/D$
as seen in Fig. \ref{FIG5}, corroborating the observation that larger gaps, influence of the half-circles on the vortex rotation.
For $D=30$, and $D=40$,the curves initially decay faster
  to zero than for smaller diameters [Figs.  \ref{FIG4}(a)-(b)]. 
  Note that the stability-transition points in Fig.~\ref{FIG5} do not correspond to zero rotation, but rather to enhanced fluctuations, indicating that the influence of the half-circles decreases.

 \section{conclusions}
 
 We studied the control
 of vortex  rotation in dry active matter by surrounding a central obstacle with $M$ half-circles. We measured the vortex angular velocity, the function $\Pi_{M}$, defined in Eq. (\ref{eq:orderparameter}) as functions of  the
  orientation angle ($\alpha$), the obstacle size $D$, and the minimum gap $\lambda$ between the half-circles and the central obstacle, when $\alpha\geq 0$. We observed two regimes depending on the value of $\alpha$. For $\alpha\geq 0$  (curved sides facing the central obstacle), the vortex has a counter-clockwise rotation ($\Pi_{M}>0$); on  the other hand, for $\alpha<0$ (flat sides facing the central obstacle), the vortex has a clock-wise rotation ($\Pi_{M}<0$). For $\alpha = \pm90^{\mathrm{o}}$, the vortex 
  did not show a net rotation. No net rotation is observed because the induced particle currents are symmetric and cancel each other, and at $\alpha = 0$, the induced currents changed direction, also resulting in no net rotation. 
 For $\lambda>0.1D$ and $M=4$, the angular velocity $\omega_{M}$ of the controlled vortex could occasionally exceed that of the isolated vortex, as shown in Fig. \ref{FIG2}(a) ($\Pi_{4}>1$) and Figs.~\ref{FIG3}(a)–(d).
 %talk about maximum and minio=mal angles
 % cite two  work of potiguar
 %not fogot to speak about  the fact angular velocity is bigger than isolatedd one

 We also investigated the effect of control on vortex stability, focusing on the 4HC configuration with $\alpha = 50$. 
 For all values of $D$, the influence of the half-circles was stronger at small gaps, i.e., when they were close to the obstacle. 
 At large gaps, comparable to the obstacle diameter, the vortex became effectively isolated. 
 By monitoring the number of reversals of the vortex rotation, we found that at small gaps the vortex was more stable, showing fewer reversals, whereas at large gaps it recovered the natural stability of an isolated vortex. 
 Between these two regimes, we identified stability-transition points that depended on both $\lambda$ and $D$.

  %According with pan and other \cite{pan2020vortex} show the relation between sizer of circular obstacle ,$D$, and stability of particle's vortex. For small size, the random vortex state; for intermediary size obstacle, the transient vortex size; and finally for  big size, vortex state or  stable size. The random vortex state exhibits more rotation changes than the transient state, while the stable vortex has very few of them.We defined the frequency of instability zones in eq. (\ref{eq:frequency}), $f$, to characterize such states. We show that random and transient vortices can be change their naturally state of stability  at small ration $\lambda/D$, see Fig. \ref{FIG5} (a) (random to transient vortex state and  transient to stable vortex state). We attribute such behavior due competition between natural fluctuations and stability imposed by half-circles setup. 

  Unexpectedly, we found that vortices that were naturally stable ($D \geq 30$) could transition to a transient regime for specific values of $\lambda / D$. 
  When the half-circles were close to the obstacle, the vortex was strongly influenced by them; 
  when they were farther away, it behaved essentially as an isolated stable vortex, rarely reversing its rotation. 
  Consequently, there existed specific values of $\lambda / D$ at which the system underwent transitions between stable and transient states.

A possible direction for future work is to extend the present setup to a regular lattice of circular obstacles and half-circles, 
in order to explore whether controlled lattice states---similar to those observed in wet models \cite{reinken2022ising,wioland2016ferromagnetic}-emerge. 
This remains a challenging problem, since achieving a uniform distribution of particles across the lattice is difficult 
due to the clustering tendency observed in dry active matter.

%The $\Pi_{M}$ function exhibits a plateau for a wide range of the $\lambda>2$; see Fig\ref{FIG1}(a), (b),(c), (d) and (e); showing noticeable change only when the $\lambda$ parameter is comparable with size of the central obstacle, $D$. We show that, our control parameter, $\Pi_{4}$, tend decrease as increase the ration $\frac{\lambda}{D}$ for all values of $D$, see Figs. \ref{FIG4}(a), (b) and (c).
 %transitions with the plateau observed in the $\Pi_4$ curve as a function of $\lambda/D$, shown in Fig. \ref{FIG4}(c). The transition points from stable to transient vortex states in the phase diagram (Fig. \ref{FIG5}(a)) are found at the end of this plateau in the $\Pi_4$ curve.

 %Our findings indicate that strong vortex stability, combined with large fluctuations, makes the control of the central vortex more challenging. Finally, we show that for intermediate obstacle sizes ($D=20$ and $25$), the vortex becomes more controllable when $\lambda/D < 0.8$. This suggests that transient vortices are more responsive to control mechanisms.

 A possible future direction for this investigation is the extension of the
 current setup to a regular lattice of circular obstacles and half-circles in order to explore
 whether any controlled lattice state, similar to those observed for wet models
 , emerges. This remains a challenging problem because achieving
 an uniform distribution of particles throughout the lattice is rather difficult,
 given the clustering tendency observed in dry active matter.

 \label{conclusions}

%\bibliography{ref}% Produces the bibliography via BibTeX.

\end{document}